\documentclass[3p,times,procedia]{elsarticle}
\usepackage{ecrc}

\usepackage[utf8]{inputenc}
\usepackage[hyphens]{url}
\usepackage{xcolor}
\usepackage{amsmath}
\usepackage{amssymb}
\usepackage{booktabs}
\usepackage{tabularx}
\usepackage{multirow}
\usepackage{graphicx}
\usepackage{tikz}
\usetikzlibrary{positioning,arrows.meta,calc}
\usepackage[bookmarks=false]{hyperref}
    \hypersetup{colorlinks, linkcolor=blue, citecolor=blue, urlcolor=blue}

\biboptions{numbers,sort&compress}

\volume{00}
\firstpage{1}
\journalname{Procedia Computer Science}
\runauth{P. Sandulu et al.}
\jid{procs}

\input{tables/numbers}

\begin{document}
\begin{frontmatter}

\dochead{8th International Conference on AI in Computational Linguistics}

\title{Premise Recovery in Software Design:\\ What Does This Decision Rest On?}

\author[a]{Priyabanta Sandulu\corref{cor1}}
\author[a,b]{Islam Mesabah}
\author[a,b]{Sebastian J. Vollmer}

\address[a]{German Research Center for Artificial Intelligence (DFKI), Trippstadter Stra{\ss}e 122, 67663 Kaiserslautern, Germany}
\address[b]{University of Kaiserslautern-Landau (RPTU), Gottlieb-Daimler-Stra{\ss}e 47, 67663 Kaiserslautern, Germany}

\begin{abstract}
%
Design decisions shape software architecture, and writing them down preserves the reasons behind it.
Architecture decision records (ADRs) capture them, each with its context, choice, and consequences.
A decision can depend on earlier ones: reversing one of them would force its review.
We call these its premises.
Yet ADRs do not always name their premises.
When a decision is reversed, engineers supersede its ADR and search by hand for what it affects.
That search can miss ADRs resting on an unnamed premise.
We frame this as premise recovery: given an ADR with identifiers masked, find its premises among earlier records.
We collect 1,060 ADRs from 36 repositories and label 1,529 candidate pairs for 60 target ADRs.
Two LLM annotators agree at $\kappa = 0.859$.
An independent human annotator, blind to their labels, agrees with the stated/implicit tag on 30 of 32 premises ($\kappa = 0.875$).
We compare 7 open-weight models with 2 baselines that use no LLM.
Qwen3.8-27B reaches F1 \res{QTD}{F1}, against \res{LEX}{F1} for word overlap; 2 models added post hoc score similarly.
On the \res{all}{nLow} premises whose titles barely overlap the target, it finds \res{QTD}{Rlow}, while word overlap finds none.
Reasoning raises its recall of implicit premises but not its F1.
Word overlap with as many predictions gains \res{all}{gainShare} as much.
We find only 2 to 3 stale dependents, though we searched for unlinked ones in only 30 of 66 supersessions.
Open models could thus suggest unwritten premises when a decision is reversed, with precision between 0.68 and 0.78.
\end{abstract}

\cortext[cor1]{Corresponding author.}
\end{frontmatter}

\email{priyabanta.sandulu@dfki.de}

\section{Introduction}
\label{sec:intro}

Software teams record design decisions in architecture decision records (ADRs), each with a context, a decision, and consequences~\cite{nygard2011adr}.
Open-source projects on GitHub use ADRs~\cite{buchgeher2023adr,desantana2026adr}, and a decision rarely stands alone.
In GOV.UK's Content Publisher\footnotemark, ADR 9 persists each edition and revision of a document to the application's database.
This presupposes ADR 2, which chose a local datastore that syncs to the central Publishing API.
If the team reversed ADR 2, ADR 9 would need review, yet ADR 9 never names ADR 2.
Finding what a change affects is change impact analysis, studied with LLMs for requirements~\cite{etezadi2025impact}; for ADRs, part of these dependencies is unwritten.
Annotators tag 16 of the 58 premises with an agreed tag as implicit: the dependent ADR does not refer to them.

\begin{figure}[t]
\centering
\includegraphics[width=0.88\linewidth]{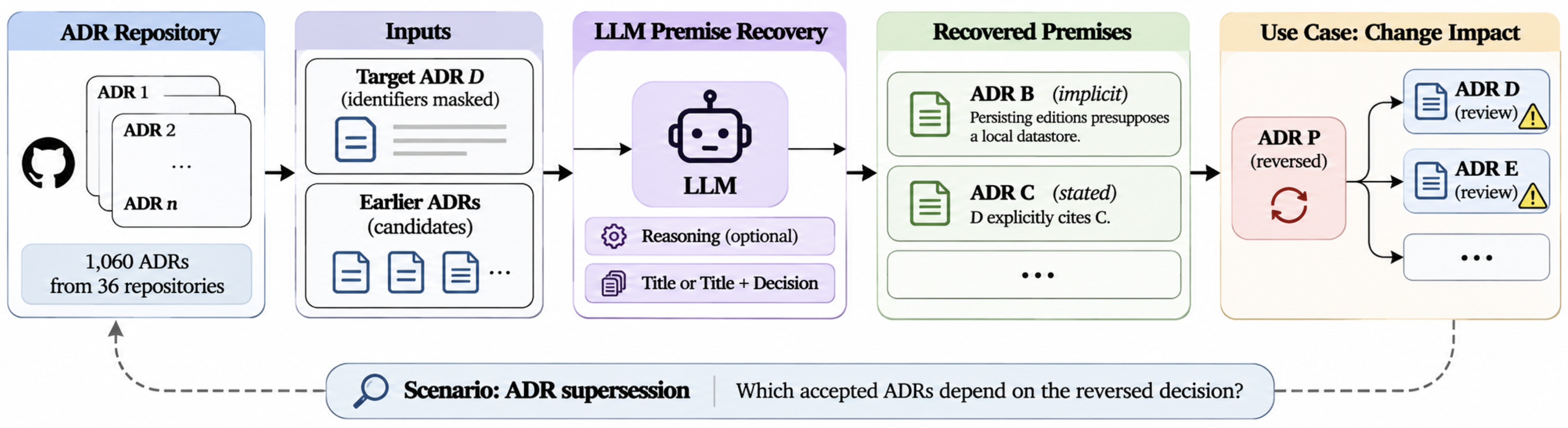}\\
\caption{Premise recovery. Inputs: a target ADR $D$ with identifiers masked and a closed list of earlier ADRs. An LLM sees titles (T) or titles and decision text (TD), with optional reasoning. It returns the premises of $D$, stated or implicit. Example from \texttt{alphagov/content-publisher} (ADR 9). Use case: reversing a premise $P$ flags its dependents for review (Section~\ref{sec:rq3}).}
\label{fig:task}
\end{figure}

An earlier decision $P$ is a \emph{premise} of $D$ if reversing $P$ would force a review of $D$.
It is \emph{stated} if $D$'s text refers to it, and \emph{implicit} otherwise.
Stated premises can be found by matching references; implicit ones require reading both decisions.
Reconstructing unstated premises is an established problem in argument mining~\cite{boltuzic2016fill,habernal2018arct,chakrabarty2021implicit}.
There, the missing premise links a claim to its support; here, it is a separate, earlier record (Figure~\ref{fig:task}).

\footnotetext{\url{https://github.com/alphagov/content-publisher/tree/a268264bd0ccfb4956f401c328bacced0b5a1d93/docs/adr}}
This paper asks 3 questions:
\begin{description}
\item[RQ1] Can language models recover premises when ADR identifiers are masked, including premises the text does not reveal?
\item[RQ2] Do decision text, model scale, and reasoning improve recovery?
\item[RQ3] How often does a superseded ADR leave a dependent that remains accepted?
\end{description}

Our contributions are:
(1) a corpus of 1,060 ADRs from 36 repositories with typed cross-reference links;
(2) an evaluation set of 60 decisions over 1,529 candidate pairs, labelled by 2 LLM annotators, with a priming check;
(3) an evaluation of 7 open-weight models and 2 baselines over 3 seeds, with repository-level intervals and corrected tests chosen after a pilot;
(4) a measurement of how the label source changes conclusions.
Code, item lists, labels, and model outputs are publicly available.\footnote{\url{https://github.com/sanduluP/adr-premise-recovery}}

\section{Related Work}
\label{sec:related}

\paragraph{Traceability and dependency detection}
Traceability link recovery connects software artifacts such as requirements and code; LiSSA performs it with retrieval-augmented generation~\cite{fuchss2025lissa}.
LLMs also detect dependencies between requirements~\cite{darif2026dependencies} and analyse the impact of requirement changes~\cite{etezadi2025impact}.
Premise recovery is link recovery between decision records; it also resembles citation attribution~\cite{press2024citeme}.
It differs in 2 ways.
A link holds only if reversing the earlier decision forces a review, and the text need not cite it.

\paragraph{Architectural decisions and LLMs}
Zimmermann et al.\ manage dependency relations in decision models~\cite{zimmermann2009dependency}; we recover them from text.
Gupta et al.\ find that a short window of earlier ADRs improves LLM-based ADR generation~\cite{gupta2026context}.
Buchgeher et al.\ and Garcia De Santana et al.\ study how open-source projects adopt ADRs~\cite{buchgeher2023adr,desantana2026adr}.
LLMs have been used to generate design decisions~\cite{dhar2024adrgen} and design rationale~\cite{zhou2025rationale}, and to detect violations of recorded decisions~\cite{su2026violations}.

\paragraph{Implicit premises in argument mining}
Boltu{\v{z}}i{\'c} and {\v{S}}najder analyse implicit premises between claims from online debates~\cite{boltuzic2016fill}.
Habernal et al.\ define the task of identifying and reconstructing the implicit warrant of an argument~\cite{habernal2018arct}.
Chakrabarty et al.\ generate implicit premises with commonsense knowledge models~\cite{chakrabarty2021implicit}.
Ku et al.\ select implicit premises by a debate between 2 LLM agents~\cite{ku2025debate}.
There, the implicit premise is a missing proposition; here, it is an earlier record of the same project.
Stahl et al.\ detect and reconstruct enthymemes in learner arguments~\cite{stahl2023gap}.
On ARCT, models exploit lexical cues instead of reasoning~\cite{niven2019probing}; our budget control tests for this artefact.

\section{Data and Annotation}
\label{sec:data}

\subsection{ADR corpus}
We started from the repository list of Buchgeher et al.~\cite{buchgeher2023adr}.
We kept repositories that store ADRs as numbered text files in a dedicated directory.
We queried 39 repositories on 31 October 2026 at a fixed commit each; 36 contained ADRs, 1,060 in total.
For each ADR we parse the title, the status, and the context, decision, and consequences sections.
Statuses are normalised to 7 classes; 572 ADRs are accepted, 59 carry the status superseded, and 283 have no parsable status.
A supersession link marks 7 more as superseded, for 66 in 29 repositories.
We order ADRs within a repository by their file number or date prefix.

\paragraph{Typed links}
A regular expression finds references to other ADRs by number or file name.
A cue-phrase rule types each reference by its sentence.
The corpus contains 509 typed links, among them 146 mention, 140 depends\_on, and 69 supersession links.
We call a depends\_on or mention link to an earlier ADR a \emph{silver premise}.
There are 247 silver-premise links (211 distinct ADR pairs) from 156 ADRs in 26 repositories.

\subsection{Premise definition}
An earlier ADR $P$ is a premise of ADR $D$ if the following test holds: if $P$ had not been made, or were reversed, $D$ would need review.
A history mention or a ``see also'' link does not pass the test.
A premise is \emph{stated} if $D$ refers to $P$ by number, title, or content, and \emph{implicit} otherwise.
The guideline gives 3 worked examples; example 1 quotes \texttt{alphagov/gsp} ADR003, a candidate for 2 targets.

\subsection{LLM annotation}
\label{sec:annotation}
Two LLMs, Claude Opus 5.5 (annotator A) and Claude Sonnet 5.5 (annotator B), labelled the data independently.
Each received the guideline as its system prompt, with no tools and no shared context.
Neither model is evaluated as a system in Section~\ref{sec:results}.
A label counts as gold only if both annotators assign it.
One author resolved the 19 disagreements and accepted 9 as premises.
This resolution serves as a check, not as the main label set (see Label sets below).
An independent human annotator tagged 32 premises blind (Section~\ref{sec:discussion}).
The 3 tasks took 200 calls on 190 items (Task A batches 10 links per call); the priming check took 120.
We ran 3 tasks:
\begin{description}
\item[Task A (link check).] For 100 sampled typed links, is the target a premise of the source?
Agreement is Cohen's $\kappa = 0.691$.
Of the 60 agreed links that satisfy the silver rule, 34 are premises (57\%).
\item[Task B (implicit dependents).] For 30 superseded ADRs from 20 repositories, which ADRs written between the superseded ADR and its successor depend on it?
The 254 candidate pairs, without superseded or rejected ADRs, give $\kappa = 0.661$; 2 items were capped at 25 candidates.
\item[Task C (premise recovery).] For 60 target ADRs, which earlier ADRs of the same repository are premises?
Targets are drawn at random (seed 20261031) from ADRs that are not rejected.
Each has 3 or more earlier ADRs and a decision or context section.
They come from 23 repositories, at most 4 and 2 per repository.
Of these, 40 have a silver premise; 20 have no outgoing link, though later ADRs cite 4 of them.
Candidates are the up to 40 nearest earlier ADRs (mean 25.5, range \mbox{3--40}), giving 1,529 pairs.
Annotators saw the first 6,000 characters of the unmasked target; 25 targets are longer.
Five silver pairs on 2 \texttt{cosmos/cosmos-sdk} targets cite the premise only after this cut.
They are in no label set, and we exclude them from scoring.
Per candidate, they saw its file name, raw title, and the first 300 characters of its decision section.
Agreement is $\kappa = 0.859$.
Both annotators mark 61 pairs: 42 stated, 16 implicit, and 3 with different tags.
The guideline's rule to leave a candidate out when unsure means more than 16 may exist.
They disagree on 19 pairs; 14 targets have no agreed premise.
\end{description}

\paragraph{Priming check}
Guideline example 3 (Swagger presupposes REST) could prime annotators toward technology presuppositions, so we re-ran Task C without it.
The new consensus has 60 pairs: 45 stated, 14 implicit, 1 disputed tag.
Of these, 57 match the original consensus.
Of the 16 original implicit premises, 12 recur; all 4 lost pairs are implicit.
We did not measure how much a repeated run with the same guideline varies.

\paragraph{Label sets}
The main labels drive every headline result (Tables~\ref{tab:main}--\ref{tab:tests}, Figure~\ref{fig:findings}, abstract).
They are the 61 consensus pairs, without the 19 disagreement and 5 late-cited pairs.
Table~\ref{tab:labels} adds 5 checks: Strict and Lenient score the 19 pairs as non-premises and as premises.
Adjudicated uses the author's resolution, Silver the 46 rule-based links; No-pred.\ drops 4 pairs.

\section{Experimental Setup}
\label{sec:setup}

\paragraph{Task input}
Each item gives the model a target ADR $D$ and the list of its candidate ADRs.
Like the annotators, the model sees the first 6,000 characters of $D$. Of the 60 masked targets, 24 are longer (masking shortens 1).
We mask ADR identifiers in $D$ and in the candidate titles, including numbers after relation cues (``Requires 46'').
One of the 120 prompts (60 targets, 2 strategies) keeps an ADR number in a candidate excerpt.
The No-pred.\ label set (Section~\ref{sec:robust}) excludes that pair with 3 others.
Candidates carry letter labels (A to AN), so a stated premise must be recognised by its title or content.
The model returns a JSON list of candidate labels with a short reason each.

\paragraph{Strategies}
\emph{Title (T)} shows candidate titles only.
\emph{Title + decision (TD)} adds the first 300 characters of each candidate's decision section, or of its context section if the decision section is missing.
Qwen3.8-27B and Gemma4-26B run with reasoning on (+R) and off; gpt-oss-20B and GLM-5.3 always reason; Qwen3-8B never.

\paragraph{Systems}
We test 7 open-weight models.
Ollama serves Qwen2.5-7B-Instruct~\cite{qwen25}, Qwen3-8B~\cite{qwen3}, and Llama-3.1-8B~\cite{llama3} on a laptop.
It serves Qwen3.8-27B~\cite{qwen38} and gpt-oss-20B~\cite{gptoss} on a shared NVIDIA H100.
Qwen3.8-27B is a hybrid-attention vision-language model; we run its 4-bit build \texttt{qwen3.8:27b-mtp-q4\_K\_M}.
gpt-oss-20B activates 3.6B of 21B parameters per token; we run it at reasoning level medium.
Each model uses its publisher's recommended sampling, except Qwen3.8-27B (Qwen3's: temperature 0.6 with reasoning, 0.7 without).
Its card recommends 1.0 with reasoning and a presence penalty of 1.5 without; reasoning effort stays at its default.
At the card's sampling, F1 changes by \res{p3}{diff} [$\res{p3}{lo}$, $\res{p3}{hi}$] for TD+R and \res{p4}{diff} for T+R (post hoc).
Two models ran post hoc.
Gemma4-26B~\cite{gemma4} activates 3.8B of 25.2B parameters; it ran on a second H100 at its card's sampling.
That is temperature 1.0, top\_p 0.95, and top\_k 64 (digest \texttt{001e5dafc3c7}).
GLM-5.3~\cite{glm53card,glm5}, a mixture-of-experts model of about 753B parameters, ran on an academic API at an unreported precision.
It reasons at its default effort (max), with temperature 1.0 and top\_p 0.95 from its card's evaluation settings.
Its reasoning is long: we reran the 73 calls that hit the 8,192-token cap at 32,768.
Still, 45 of its 360 calls reach the cap and count as empty; on its complete calls, TD+R reaches F1 0.835.
At 8,192, 4 of the other models' 3,240 calls were cut, and count as empty.
We run 3 seeds per configuration; Ollama models use a 16,384-token context.
The 3,600 LLM calls read 8.6M prompt tokens; reasoning adds 3,800 to 8,300 characters per call (GLM-5.3: 43,000).
Two baselines need no LLM.
\emph{Lexical} predicts a candidate if the share of its title words that occur in $D$ exceeds a threshold.
\emph{bge-m3} predicts a candidate if the cosine similarity of M3 embeddings~\cite{chen2024m3} exceeds a threshold.
Both thresholds (0.857 and 0.625) are tuned on silver premises of 116 ADRs outside the 60 targets.
Tuned on gold labels by leave-one-repository-out instead, they reach F1 0.446 and 0.381 (upper bound 0.489, 0.462).
Lexical reads the full masked target, and bge-m3 its first 4,000 characters.

\paragraph{Metrics}
We report pair-level precision, recall, and F1 against the consensus labels, averaged over seeds.
The 60 targets are 5.7\% of the corpus, and the 61 gold premises 4.0\% of the pairs. One premise is 0.016 of recall.
We also split recall without the stated/implicit tag.
A premise has \emph{low overlap} if under half its title words occur in the masked $D$.
This holds for \res{all}{nLow} of 61 premises, \res{all}{nImLo} tagged implicit and 4 stated ones (2 cited by number, 2 described by content).
Disagreement and late-cited pairs are excluded from scoring (the main label set); Section~\ref{sec:robust} bounds every possible resolution.
95\% confidence intervals come from a 2-level bootstrap with 2,000 resamples.
It resamples the 23 repositories, since their targets are not independent, and then the seeds of each system.
We test 9 comparisons with a paired version of this bootstrap and Holm's correction over all 9.
The $p$-value is twice the smaller share of resampled differences on either side of 0, with a floor of 1/2,000.

\section{Results}
\label{sec:results}

\begin{figure}[t]
\centering
\begin{minipage}[b]{0.46\textwidth}\centering
\includegraphics[width=0.92\linewidth]{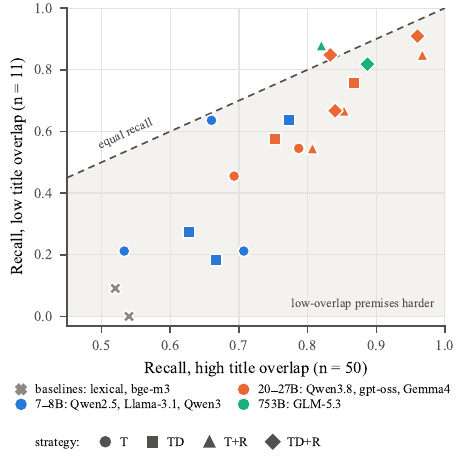}\\
{\small (a) Recall on high- vs.\ low-overlap premises}
\end{minipage}\hfill
\begin{minipage}[b]{0.52\textwidth}\centering
\footnotesize
\setlength{\tabcolsep}{3.5pt}
\begin{tabular}{lccccc}
\toprule
 & & \multicolumn{2}{c}{F1} & \multicolumn{2}{c}{R\textsubscript{im}} \\
\cmidrule(lr){3-4}\cmidrule(lr){5-6}
System & Pred & LLM & Lex. & LLM & Lex. \\
\midrule
Qwen3.8-27B TD & 1.47 & 0.756 & 0.511 & 0.667 & 0.333 \\
Qwen3.8-27B TD+R & 1.82 & 0.750 & 0.502 & 0.938 & 0.521 \\
Qwen3.8-27B T & 1.24 & 0.706 & 0.509 & 0.667 & 0.333 \\
Qwen3.8-27B T+R & 2.10 & 0.666 & 0.510 & 0.875 & 0.521 \\
gpt-oss-20B TD+R & 1.85 & 0.624 & 0.509 & 0.729 & 0.521 \\
Qwen3-8B T & 1.23 & 0.586 & 0.412 & 0.229 & 0.312 \\
\bottomrule
\end{tabular}
\\[6pt]
{\small (b) 6 LLM configurations vs.\ word overlap with the same number of predictions per target}
\vspace{4pt}
\end{minipage}
\caption{(a) Main labels split by title overlap with the masked target (no stated/implicit tag); colour: model size, marker: strategy (legend below); lexical is the lower grey cross; x-axis starts at 0.45. (b) Lex.: the lexical scorer predicts, per target, as many top-scoring candidates as the LLM did; Pred counts all pairs, including the 19 disagreement and 5 late-cited pairs. Means over seeds.}
\label{fig:findings}
\end{figure}

\begin{table}[t]
\caption{Premise recovery on 60 targets (1,529 pairs; 61 gold premises; 19 disagreement and 5 late-cited pairs excluded; Gemma4-26B T, T+R in text). Str.: T titles, TD + decision text, +R reasoning. Means and SD over 3 seeds; CI: 2-level bootstrap. Res.\ range: F1 with the 19 disagreement pairs all against or all for it. R\textsubscript{lo}, R\textsubscript{st}, R\textsubscript{im}: recall on 11 low-overlap, 42 stated, and 16 implicit premises. Pred: predictions per target (gold 1.02).}
\label{tab:main}
\centering
\footnotesize
\setlength{\tabcolsep}{4pt}
\begin{tabular}{llcccccccc}
\toprule
Model & Str. & P & R & F1 (SD) [95\% CI] & Res.\ range & R\textsubscript{lo} & R\textsubscript{st} & R\textsubscript{im} & Pred \\
\midrule
Lexical & -- & 0.491 & 0.443 & 0.466 [0.326, 0.589] & 0.400--0.508 & 0.000 & 0.571 & 0.125 & 0.92 \\
bge-m3 & -- & 0.429 & 0.443 & 0.435 [0.322, 0.541] & 0.378--0.478 & 0.091 & 0.548 & 0.125 & 1.05 \\
\midrule
Qwen2.5-7B & T & 0.471 & 0.475 & 0.473 (0.025) [0.379, 0.571] & 0.410--0.495 & 0.212 & 0.563 & 0.333 & 1.03 \\
Qwen2.5-7B & TD & 0.257 & 0.563 & 0.350 (0.039) [0.242, 0.500] & 0.319--0.398 & 0.273 & 0.690 & 0.312 & 2.27 \\
Llama-3.1-8B & T & 0.150 & 0.656 & 0.243 (0.020) [0.165, 0.347] & 0.230--0.276 & 0.636 & 0.698 & 0.583 & 4.51 \\
Llama-3.1-8B & TD & 0.091 & 0.749 & 0.163 (0.035) [0.102, 0.263] & 0.157--0.195 & 0.636 & 0.770 & 0.750 & 8.52 \\
Qwen3-8B & T & 0.557 & 0.617 & 0.586 (0.009) [0.488, 0.682] & 0.510--0.621 & 0.212 & 0.770 & 0.229 & 1.13 \\
Qwen3-8B & TD & 0.381 & 0.579 & 0.458 (0.037) [0.339, 0.603] & 0.408--0.495 & 0.182 & 0.683 & 0.292 & 1.57 \\
\midrule
gpt-oss-20B & T+R & 0.411 & 0.820 & 0.547 (0.005) [0.468, 0.636] & 0.496--0.581 & 0.667 & 0.881 & 0.667 & 2.03 \\
gpt-oss-20B & TD+R & 0.499 & 0.836 & 0.624 (0.030) [0.533, 0.721] & 0.559--0.657 & 0.848 & 0.865 & 0.729 & 1.71 \\
\midrule
Qwen3.8-27B & T & 0.674 & 0.743 & 0.706 (0.027) [0.603, 0.807] & 0.615--0.731 & 0.545 & 0.770 & 0.667 & 1.12 \\
Qwen3.8-27B & TD & 0.683 & 0.847 & 0.756 (0.016) [0.697, 0.812] & 0.664--0.793 & 0.758 & 0.929 & 0.667 & 1.26 \\
Qwen3.8-27B & T+R & 0.514 & 0.945 & 0.666 (0.016) [0.585, 0.763] & 0.600--0.708 & 0.848 & 0.968 & 0.875 & 1.87 \\
Qwen3.8-27B & TD+R & 0.620 & 0.951 & 0.750 (0.023) [0.661, 0.847] & 0.668--0.791 & 0.909 & 0.952 & 0.938 & 1.56 \\
\midrule
Gemma4-26B & TD & 0.791 & 0.721 & 0.754 (0.024) [0.652, 0.851] & 0.649--0.782 & 0.576 & 0.817 & 0.417 & 0.93 \\
Gemma4-26B & TD+R & 0.783 & 0.809 & 0.796 (0.005) [0.721, 0.868] & 0.690--0.816 & 0.667 & 0.849 & 0.667 & 1.05 \\
\midrule
GLM-5.3 & T+R & 0.643 & 0.831 & 0.724 (0.035) [0.636, 0.801] & 0.637--0.762 & 0.879 & 0.849 & 0.833 & 1.32 \\
GLM-5.3 & TD+R & 0.738 & 0.874 & \textbf{0.799} (0.029) [0.702, 0.885] & 0.700--0.825 & 0.818 & 0.857 & 0.896 & 1.21 \\
\bottomrule
\end{tabular}

\end{table}

\subsection{RQ1: Recovery of stated and implicit premises}
Both baselines find about half of the stated premises (recall \res{LEX}{Rst} and \res{BGE}{Rst}), but only 2 of 16 implicit ones.
The highest F1 values are GLM-5.3 TD+R (\res{ZTDR}{F1} [\res{ZTDR}{Flo}, \res{ZTDR}{Fhi}]) and Gemma4-26B TD+R (\res{MTDR}{F1}).
The best pre-planned system is Qwen3.8-27B TD (\res{QTD}{F1}).
Neither leader differs significantly from it (post hoc: $\Delta$ \res{p5}{diff} [$\res{p5}{lo}$, $\res{p5}{hi}$] and \res{p2}{diff} [$\res{p2}{lo}$, $\res{p2}{hi}$]).
Qwen3.8-27B TD is \res{c5}{absdiff} above bge-m3 (Table~\ref{tab:tests}, c5) and \res{p1}{absdiff} above the lexical baseline (p1, post hoc).
Low-overlap recall is below high-overlap recall for \res{all}{belowDiagOv} of the \res{all}{nLLM} LLM configurations (Figure~\ref{fig:findings}a).
The best 7--8B configuration, Qwen3-8B T, beats bge-m3 after correction (F1 \res{ET}{F1}; c6).
Unparsed outputs (\res{all}{parseErr}, 49 of them cut at the output cap) count as empty; \res{all}{invalidAll} out-of-list labels are ignored.

Of the \res{all}{nLow} low-overlap premises, lexical finds none (its threshold 0.857 exceeds 0.5); bge-m3 finds \res{BGE}{Rlow}.
Qwen3.8-27B TD finds \res{QTD}{Rlow} [\res{QTD}{RlowLo}, \res{QTD}{RlowHi}], \res{QTD}{Rlown} of \res{all}{nLow} on average.
TD+R finds \res{QTDR}{Rlown}, GLM-5.3 TD+R \res{ZTDR}{Rlown}, gpt-oss-20B TD+R \res{GTDR}{Rlown}, and Gemma4-26B TD+R \res{MTDR}{Rlown}.
The \res{all}{nImLo} tagged implicit among them involve 5 premise ADRs in 4 repositories; 3 share content-publisher ADR 2.
Under the LLM tags, implicit recall is lower than stated recall for \res{all}{belowDiag} of the \res{all}{nLLM} configurations.
At Figure~\ref{fig:findings}b's budget, Qwen3.8-27B TD still leads in F1 (\res{QTD}{F1} vs.\ \res{QTD}{bmlexF1}) and implicit recall (\res{QTD}{Rim} vs.\ \res{QTD}{bmlexRim}).

\subsection{RQ2: What helps}
\paragraph{Decision text}
For Qwen3.8-27B, decision text changes F1 by \res{c1}{diff} (Table~\ref{tab:tests}, c1), and the interval includes 0.
For the smaller models, TD lowers F1 (Qwen3-8B: \res{ET}{F1} to \res{ETD}{F1}) as they predict more.

\paragraph{Scale}
Qwen3.8-27B T exceeds Qwen3-8B T by \res{c4}{absdiff} F1 (c4).
The uncorrected interval excludes 0; after Holm's correction, $p = \res{c4}{pholm}$.
The 2 models also differ in generation and architecture, so c4 is not a clean scale contrast.

\paragraph{Reasoning}
Reasoning does not raise the F1 of Qwen3.8-27B (c2: \res{c2}{diff}; c3: \res{c3}{diff}).
With TD, it raises Qwen3.8-27B's implicit recall from \res{QTD}{Rim} to \res{QTDR}{Rim}, and precision falls from \res{QTD}{P} to \res{QTDR}{P}.
For Gemma4-26B, reasoning raises F1 by \res{p7}{diff} [$\res{p7}{lo}$, $\res{p7}{hi}$] with titles (post hoc, uncorrected $p = \res{p7}{p}$).
With TD, its gain of \res{p6}{diff} [$\res{p6}{lo}$, $\res{p6}{hi}$] is not significant.
The implicit gain is not significant after correction, with or without decision text (c8: $p_\mathrm{Holm} = \res{c8}{pholm}$; c9).
Reasoning also predicts more (Figure~\ref{fig:findings}b). At the same budgets, word overlap raises implicit recall from \res{QTD}{bmlexRim} to \res{QTDR}{bmlexRim}, \res{all}{gainShare} of the model's gain of \res{all}{gainLLM}.

\begin{table}[t]
\caption{Paired comparisons, main labels. $\Delta$ = A $-$ B; CI and $p$: 2,000 bootstrap resamples; $p_\mathrm{Holm}$ over c1--c9 (set after a pilot); p1 is post hoc.}
\label{tab:tests}
\centering
\footnotesize
\setlength{\tabcolsep}{4pt}
\begin{tabular}{lllccc}
\toprule
ID & Comparison (A vs.\ B) & Metric & $\Delta$ [95\% CI] & $p$ & $p_\mathrm{Holm}$ \\
\midrule
c1 & Qwen3.8-27B: TD vs.\ T & F1 & $+0.050$ [$-0.027$, $+0.127$] & 0.215 & 0.645 \\
c2 & Qwen3.8-27B: TD+R vs.\ TD & F1 & $-0.006$ [$-0.081$, $+0.076$] & 0.898 & 0.898 \\
c3 & Qwen3.8-27B: T+R vs.\ T & F1 & $-0.040$ [$-0.132$, $+0.040$] & 0.324 & 0.648 \\
c4 & Qwen3.8-27B T vs.\ Qwen3-8B T & F1 & $+0.120$ [$+0.004$, $+0.247$] & 0.043 & 0.215 \\
c5 & Qwen3.8-27B TD vs.\ bge-m3 & F1 & $+0.321$ [$+0.214$, $+0.425$] & $<$0.001 & 0.005 \\
c6 & Qwen3-8B T vs.\ bge-m3 & F1 & $+0.151$ [$+0.061$, $+0.246$] & 0.001 & 0.008 \\
c7 & Qwen3.8-27B TD+R vs.\ gpt-oss-20B TD+R & F1 & $+0.126$ [$+0.055$, $+0.204$] & 0.001 & 0.008 \\
c8 & Qwen3.8-27B: TD+R vs.\ TD & R\textsubscript{im} & $+0.271$ [$+0.095$, $+0.439$] & 0.011 & 0.066 \\
c9 & Qwen3.8-27B: T+R vs.\ T & R\textsubscript{im} & $+0.208$ [$+0.000$, $+0.375$] & 0.052 & 0.215 \\
p1 & Qwen3.8-27B TD vs.\ Lexical & F1 & $+0.290$ [$+0.172$, $+0.430$] & $<$0.001 & post hoc \\
\bottomrule
\end{tabular}

\end{table}

\begin{table}[t]
\caption{F1 on 5 label sets and on main labels for 40 targets with a silver premise (Linked) and 20 without (No link). Silver: 46 rule-based links; Strict, Lenient: 19 disagreement pairs as non-premises, as premises; Main: consensus minus them; Adjud.: author's resolution (9 premises); No-pred.: Main minus 4 superseded predecessors.}
\label{tab:labels}
\centering
\footnotesize
\setlength{\tabcolsep}{4pt}
\begin{tabular}{lcccccc@{\hspace{16pt}}cc}
\toprule
System & Silver & Strict & Main & Adjud. & Lenient & No-pred. & Linked & No link \\
\midrule
Lexical & 0.528 & 0.446 & 0.466 & 0.431 & 0.457 & 0.455 & 0.515 & 0.133 \\
Qwen3-8B T & 0.563 & 0.560 & 0.586 & 0.566 & 0.568 & 0.558 & 0.704 & 0.254 \\
Qwen3.8-27B TD & 0.563 & 0.695 & 0.756 & 0.719 & 0.759 & 0.741 & 0.788 & 0.614 \\
Gemma4-26B TD+R & 0.623 & 0.755 & 0.796 & 0.740 & 0.748 & 0.782 & 0.825 & 0.676 \\
GLM-5.3 TD+R & 0.568 & 0.745 & 0.799 & 0.749 & 0.777 & 0.787 & 0.834 & 0.675 \\
\bottomrule
\end{tabular}

\end{table}

\paragraph{Model family}
gpt-oss-20B TD+R reaches F1 \res{GTDR}{F1}, \res{c7}{absdiff} below Qwen3.8-27B TD+R (c7, $p_\mathrm{Holm} = \res{c7}{pholm}$).

\subsection{Robustness to the labels}
\label{sec:robust}
Under the worst resolution of the 19 disagreement pairs (Table~\ref{tab:main}), Qwen3.8-27B TD and TD+R keep F1 \res{QTD}{AdjMin} and \res{QTDR}{AdjMin}.
That exceeds the best resolution for every 7--8B system (\res{all}{maxSmallAdj}) and both baselines (\res{all}{maxBaseAdj}).
Under the author's resolution (Adjud., Table~\ref{tab:labels}), Qwen3.8-27B TD reaches F1 \res{QTD}{Fadjudicated}.
Counting the 3 gold misses from the error check as premises raises Qwen3.8-27B TD's precision from 0.683 to 0.700 (pooled over seeds).
Against silver links (Table~\ref{tab:labels}), the lexical baseline (\res{LEX}{Fsilver}) ranks \res{LEX}{ranksilverord} of the \res{all}{nSys} systems. Against the main labels, it ranks \res{LEX}{rankmainord}.
Above it against silver are the 4 Gemma4-26B configurations, GLM-5.3 TD+R, Qwen3-8B T, and Qwen3.8-27B TD.

\subsection{RQ3: Stale dependents}
\label{sec:rq3}
Of 66 superseded ADRs in 29 repositories, silver links connect 6 dependents; 1 accepted one is stale (\texttt{nulib/meadow} 0008).
Task B covers 30 of the 43 supersessions with a typed successor and a candidate.
Both annotators flag \texttt{nulib/meadow} 0018, whose storage keys use a format ADR 0023 replaced; 1 annotator flags 1 more.
More may exist, as Task B omits dependents written after the successor, but far below our pre-set 20 in 3 repositories.

\section{Discussion and Limitations}
\label{sec:discussion}

\paragraph{LLM labels}
The 2 Claude annotators come from 1 model family and may share a bias.
LLM evaluators also favour their own generations~\cite{panickssery2024selfpref}.
Annotators from another model family or a panel of experts would test this bias.
On all 16 implicit and 16 sampled stated premises, an independent annotator agrees blind with the LLM tag on 30.
That is $\kappa = 0.875$ (95\% CI $[0.69, 1.00]$); both differences are LLM-implicit pairs.

\paragraph{Errors}
False positives are 32\% of Qwen3.8-27B TD's 227 predictions, 26\% of GLM-5.3 TD+R's, and 22\% of Gemma4-26B TD+R's.
The author judged Qwen3.8-27B TD's 46 distinct wrong pairs blind, mixed with 14 correct ones (13 confirmed).
Of the 46, 32 are same-topic, 6 unrelated, 4 superseded by $D$, 1 history, and 3 gold misses.
This check covers only negatives that a model predicted; a random sample of non-premise pairs remains unchecked.
52 of its 72 false-positive predictions are same-topic.
GLM-5.3 and Gemma4-26B repeat 11 and 10 of these pairs (6 and 7 same-topic); their other 14 and 12 are unchecked.
GLM-5.3 predicts 2 of the 3 gold misses, Gemma4-26B 1.

\paragraph{Contamination and models}
The ADRs are public and may be in the models' training data; masking hides identifiers, not memorised content.
We fixed the pipeline and 9 comparisons after a pilot on the same items; a 10th (Qwen3-8B with reasoning) was not run.
All analyses beyond c1--c9 are post hoc, including Gemma4-26B and GLM-5.3; annotator rerun variance is unmeasured.
Masking makes stated premises harder; the closed list spares retrieval.

\section{Conclusion}
\label{sec:conclusion}
GLM-5.3, Gemma4-26B, and Qwen3.8-27B reach F1 \res{ZTDR}{F1}, \res{MTDR}{F1}, and \res{QTD}{F1}, far ahead of word overlap.
On the \res{all}{nLow} premises with low title overlap, word overlap finds none and Qwen3.8-27B \res{QTD}{Rlow}.
With titles, reasoning raises Gemma4-26B's F1 (post hoc); for Qwen3.8-27B, reasoning, decision text, and scale do not after correction (c1--c4).
\paragraph{Future work}
Retrieval over all earlier ADRs would test the full task.
ADRs written after training cut-offs would rule out memorisation.
Asking whether reversing each candidate forces review could cut same-topic false positives.
More human annotators, other reasoning efforts, and Qwen3-8B with reasoning would complete the comparisons.
A tool could flag likely dependents at supersession, tested with practitioners.

\section*{Declaration of generative AI and AI-assisted technologies in the manuscript preparation process}
Claude Opus 5.5 and Sonnet 5.5 annotated the data.
Claude Code assisted with code, drafting, and review.
GPT (OpenAI) drew Figure 1 from the authors' outline.
The authors checked all outputs and take full responsibility.

\bibliographystyle{elsarticle-num}
\bibliography{references}

\end{document}